\documentclass[prl,twocolumn,amsfonts,amsmath,tightenlines,preprintnumbers,nofootinbib,superscriptaddress,letterpaper]{revtex4}
\usepackage{graphicx}
\usepackage{bm}

\newcommand{\Choose}[2]{{\begin{pmatrix} {#1} \\ {#2} \end{pmatrix}}}

\newcommand{\lsim}{\raisebox{-0.7ex}{$\stackrel{\textstyle <}{\sim}$ }}

\begin{document}

\preprint{NT@UW-07-15}
\preprint{UNH-07-04}
\preprint{UCRL-JRNL-235354}
\preprint{JLAB-THY-07-735}

\begin{figure}[!t]
\vskip -1.1cm
\leftline{
\includegraphics[width=2.0 cm]{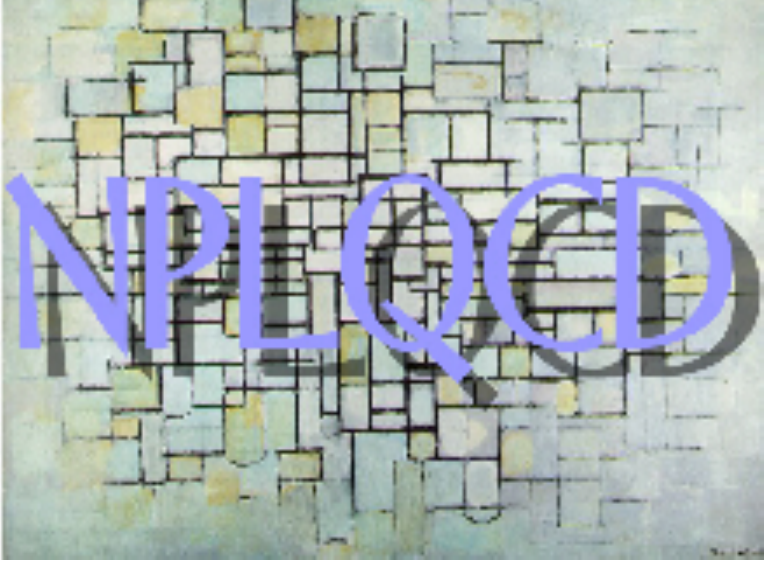}}
\vskip -0.5cm
\end{figure}

\title{Multi-Pion Systems in  Lattice QCD  and the Three-Pion Interaction
}

\author{Silas R. Beane} \affiliation{Department of Physics, University
  of New Hampshire, Durham, NH 03824-3568, USA}

\author{William Detmold} \affiliation{Department of Physics,
  University of Washington, Box 351560, Seattle, WA 98195, USA}

\author{Thomas C.~Luu}
\affiliation{N Division, Lawrence Livermore National Laboratory, Livermore, CA
  94551, USA.}
\author{Kostas Orginos}
\affiliation{Department of Physics, College of William and Mary, Williamsburg,
  VA 23187-8795, USA}
\affiliation{Jefferson Laboratory, 12000 Jefferson Avenue, 
Newport News, VA 23606, USA.}

\author{Martin J. Savage} \affiliation{Department of Physics,
  University of Washington, Box 351560, Seattle, WA 98195, USA}

\author{Aaron Torok} \affiliation{ Department of Physics, University
  of New Hampshire, Durham, NH 03824-3568, USA}
\collaboration{ NPLQCD Collaboration }

\date\today

\begin{abstract}
The ground-state energies of 2, 3, 4 and 5 $\pi^+$'s in a spatial
volume $V\sim (2.5~{\rm fm})^3$ are computed with lattice QCD.  By
eliminating the leading contribution from three-$\pi^+$ interactions,
particular combinations of these $n$-$\pi^+$ ground-state energies
provide precise extractions of the $\pi^+\pi^+$ scattering length 
in agreement with that obtained from calculations involving only
two $\pi^+$'s.  The three-$\pi^+$ interaction can be isolated by
forming other combinations of the $n$-$\pi^+$ ground-state energies.
We find a result that is consistent with a repulsive three-$\pi^+$
interaction for $m_\pi\lsim 352~{\rm MeV}$.
\end{abstract}

\maketitle


A major goal of strong-interaction physics is to determine the
spectrum and interactions of hadrons and nuclei from Quantum
Chromodynamics (QCD).  Lattice QCD is the only known way to rigorously
compute strong-interaction quantities, and an increasing effort is
being put into understanding the lattice QCD calculations that will be
required to extract even the most basic properties of light nuclei.
It is clear that at some level, the interactions among three or more
hadrons play a significant role in nuclei, and an important goal for
lattice practitioners is to determine the parameters of these
interactions.  We report on the first lattice QCD calculation of
systems comprised of more than two hadrons.

The simplest multi-hadron systems (both conceptually, and from a
numerical perspective) consist of $n$ pseudoscalar mesons of maximal
isospin.  Interactions among multiple pions are important to explore
for phenomenological reasons. Two- and three-pion interferometry is
currently being used to determine the coherence of the pion source in
heavy-ion collisions~\cite{Aggarwal:2000ex}.  Further, multi-pion
interactions impact the formation of a pion condensate which is
energetically favored in systems with large isospin chemical
potential, and will influence the properties of (hot) pion gases.  In
this work we perform lattice QCD calculations of the ground-state
energies of $\pi^+\pi^+$, $\pi^+\pi^+\pi^+$, $\pi^+\pi^+\pi^+\pi^+$,
and $\pi^+\pi^+\pi^+\pi^+\pi^+$ systems in a spatial volume of $V\sim
(2.5~{\rm fm})^3$ with periodic boundary conditions and a lattice
spacing of $b\sim 0.125~{\rm fm}$.  These systems provide an ideal
laboratory for investigating multi-particle interactions, as chiral
symmetry guarantees relatively-weak interactions among pions, and
multiple-pion correlation functions computed with lattice QCD do not
suffer from signal-to-noise issues that are expected to plague
analogous calculations in multi-baryon systems.  The $\pi^+\pi^+$
scattering length is extracted from the $n>2$ pion systems with
precision that is comparable to (and in some cases better than) the
$n=2$ determination~\cite{Beane:2007xs}. Additionally, a result that
is consistent with a repulsive three-pion interaction of magnitude
expected from naive dimensional analysis (NDA) is found for
$m_\pi\lsim 352~{\rm MeV}$.

At finite volume, the ground state energy of a system of $n$ bosons of
mass $M$ is shifted from its infinite volume value, $n\,M$.  In a
periodic cubic spatial volume of periodicity $L$, this shift is known
to be~\cite{Huang:1957im,Lee:1957,Wu:1959,Luscher:1986pf,
  Luscher:1990ux,Beane:2007qr,Tan:2007bg}
\begin{eqnarray}
\label{eq:1}
  {\Delta E}_n &\!\!=&\!\!
  \frac{4\pi\, a}{M\,L^3}\Choose{n}{2}\Bigg\{1
-\frac{a\,{\cal I}}{\pi\,L}
+\left(\frac{a}{\pi\,L}\right)^2\left[{\cal I}^2+(2n-5){\cal J}\right]
\nonumber
\\&&\hspace*{-0.9cm}
-
\left(\frac{a}{\pi\,L}\right)^3\Big[{\cal I}^3 + (2 n-7)
  {\cal I}{\cal J} + \left(5 n^2-41 n+63\right){\cal K}
\Big]
\Bigg\}
\nonumber
\\&&\hspace*{-0.7cm}
+\Choose{n}{2} \frac{8\pi^2 a^3}{M\, L^6}r
+\Choose{n}{3}\frac{\overline\eta_3^{L}}{  L^6} 
\ +\  {\cal O}\left(1/L^{7}\right)
\ ,
\end{eqnarray}
where $a$ and $r$ are the two-boson scattering length and the
effective range parameter, respectively, and $\overline{\eta}_3^{L}$
is the renormalization-group invariant (RGI) three-boson interaction
($\overline{\eta}_3^{L}$ is renormalization-scheme and scale
independent, but depends logarithmically on $L$; in terms of the
three-particle interaction defined in Ref \cite{Beane:2007qr},
$\overline{\eta}_3^{L} = \eta_3(\mu)+\frac{64\pi
  a^4}{M}(3\sqrt{3}-4\pi)\log(\mu\,L) - \frac{96\,a^4}{\pi^2 M}[2
{\cal Q}+{\cal R}]$). 
The geometric constants appearing in Eq.~(\ref{eq:1}) are 
${\cal I}= -8.9136329$,
${\cal J} = 16.532316$ and ${\cal K} = 8.4019240$. At this order, the
energy is only sensitive to a combination of the effective range and
scattering length, $\overline{a}=a+\frac{2\pi}{L^3}a^3\,r$ and in what
follows we replace $a\to\overline{a}$, eliminating $r$. The above
expansion is valid provided $a,\,r \ll L$ with an additional constraint
on $n$~\cite{Beane:2007qr}.

Various combinations of the energy differences defined in
Eq.~(\ref{eq:1}) are particularly useful in what follows.  First
\begin{eqnarray}
  \label{eq:2}
  \frac{L^3 M ({\Delta E}_n m (m^2-3m+2) - {\Delta E}_m n
    (n^2-3n+2))}{2 (m-1) m (m - 
    n) (n-1) n \pi} \hspace*{2mm}
\nonumber
\\
=\overline{a} \Bigg\{
1- \frac{\overline{a}}{\pi \,L}{\cal I}
 +\left(\frac{\overline{a}}{\pi \,L}\right)^2\left[{\cal I}^2-{\cal J}\right]
+\left(\frac{\overline{a}}{\pi \,L}\right)^3
\hspace*{7mm}
\nonumber
 \\
\times\left[-{\cal I}^3 + 3
   {\cal IJ} +  (19 + 5 m \,n - 10 (n+m)){\cal K}\right]
\Bigg\}
\ ,
\hspace*{2mm}
\end{eqnarray}
(for $n,m>2$) is independent of $\overline{\eta}_3^{L}$ and allows a
determination of $\overline{a}$ up to ${\cal O}(1/L^{4})$ corrections
(combinations achieving the same result using all of the $n=3,4,5$
energies can also be constructed).  Second, the three body parameter
can be directly determined from
\begin{eqnarray}
  \label{eq:3}
  \overline{\eta}_3^{L}\!\! &=&\!\!  L^6\Choose{n}{3}^{-1}\Bigg\{{\Delta E}_n -
    \Choose{n}{2}{\Delta E}_2
    -6\Choose{n}{3} M^2 {\Delta E}_2^3
\\
&\times& \left(\frac{L}{2\pi}\right)^4
\left[ {\cal J}
      + \frac{L^2 M\,{\Delta E}_2}{2\pi^2} \left(
      {\cal IJ} -(5n-31){\cal K}\right) 
      \right]\Bigg\}
\ ,
\nonumber
\end{eqnarray}
($n>2$) with corrections arising at ${\cal O}(1/L)$.  Additionally,
the dimensionless quantity
\begin{eqnarray}
  \label{eq:4}
1 - \frac{2}{3}\frac{{\Delta E}_3}{\Delta E_2}
 +\frac{1}{6}\frac{{\Delta E}_4}{\Delta E_2} + \frac{5M^3\,L^6{\cal 
      K}}{32\pi^6}{\Delta E}_2^3  
\ \sim \ {\cal O}(1/L^7)
\,,
\end{eqnarray}
provides a useful check of the convergence of the expansion.
Combinations involving ${\Delta E}_{2,3,5}$ and ${\Delta E}_{2,4,5}$
that vanish at this order can also be constructed.

The requisite ground-state energies are extracted from 
the $n$-$\pi^+$ correlation functions defined by
\begin{eqnarray}
  \label{eq:5}
  C_n(t) &=& \Big\langle 0 \Big| \Big[\sum_{\bf x} \chi_{\pi^+}({\bf
      x},t) \overline\chi_{\pi^+}(0,0)\Big]^n \Big|0\Big\rangle
\ \ ,
\end{eqnarray}
where $\chi_{\pi^+}(x)=u^a (x) \gamma_5 \overline{d}_a(x)$ is an
interpolating operator for the $\pi^+$ ($a$ is a color index).  The
sums in Eq.~(\ref{eq:5}) project the correlation functions onto the
$A_1^+$ representation of the cubic symmetry group (in the continuum
this corresponds to angular momentum $\ell=0,4,\ldots$). As $n$
increases, the number of Wick contractions involved in computing
$C_n(t)$ increases as $ n!^2$.  In the limit of isospin symmetry, the
correlation functions in Eq.~(\ref{eq:5}) with $n<13$ require the
computation of only a single quark propagator, $S(x;0)$ (for $n>12$
additional propagators are required to circumvent the Pauli
exclusion).  As an example, the $n=3$ correlator can be expressed as
\begin{eqnarray}
  \label{eq:6}
  C_3(t) &=&
{\rm tr}\left[\Pi\right]^3 \ -\ 3\ {\rm tr}\left[\Pi\right]{\rm
  tr}\left[\Pi^2\right]
 +\ 2\ {\rm tr}\left[\Pi^3\right]\,,
\end{eqnarray}
where $\Pi=\sum_{\bf x} \gamma_5 S({\bf x},t;0)\gamma_5 S^\dagger({\bf
  x},t;0)$ and the trace is over Dirac and color indices.

In this work we have computed $C_{1,2,3,4,5}(t)$ in mixed-action
lattice QCD, using domain-wall valence quark propagators from Gaussian
smeared-sources on the rooted-staggered coarse MILC
gauge-configurations ($20^3\times 64$) after HYP-smearing and chopping
(see Refs.~\cite{Aubin:2004fs,Beane:2007xs} for details). These are
computed at pion masses of $m_\pi\sim~291,\, 352,\, 491,\, 591~{\rm
MeV}$.  Details of the propagators used in the correlation functions
are given in Table~\ref{tab:params} and can be found in
Ref.~\cite{Beane:2007xs}.
\begin{table}[!hb]
  \centering
  \begin{ruledtabular}
  \begin{tabular}{c|cc|c}
 $m_\pi$ (MeV) & $N_{\rm cfg}$ & $N_{\rm src}$ & $m_\pi/f_\pi$ \\ \hline
$291.3(1.0)(1.0)$ & 468 & 16 & 1.990(11)(14)
\\
$351.9(0.5)(0.2)$ & 769 & 20 &  2.3230(57)(30)
\\
$491.4(0.4)(0.3)$ & 486 & 24 &  3.0585(49)(95)
\\
$590.5(0.8)(0.2)$ & 564 & 8  &  3.4758(98)(60)
  \end{tabular}
  \end{ruledtabular}
  \caption{Parameters of the domain-wall propagators used herein. 
A lattice spacing of $b=0.125~{\rm fm}$ has been used 
to convert from lattice to physical units.
The number of gauge configurations is $N_{\rm cfg}$, and the number of sources
per configuration is $N_{\rm src}$.
For further details see Ref.~\protect\cite{Beane:2007xs}.
}
  \label{tab:params}
\end{table}

The energies of $n$ pion states are dominated by the $n$ single-pion
energies, with the interactions contributing a small fraction of the
total energy.  To extract the resulting energy shifts, $\Delta E_n$,
the ratios of correlators
\begin{equation}
  \label{eq:7}
  G_n(t)= \frac{C_n(t)}{\left[C_1(t)\right]^n}
  \stackrel{t\to\infty}{\longrightarrow}
 A e^{-\Delta E_n \,t}\,,
\end{equation}
are formed, where the second relation holds in the limit of infinite
temporal extent and infinite number of gauge configurations.
Inclusion of the effects of temporal boundaries (here Dirichlet
boundary conditions are used) is complicated for multi-hadron systems,
and our analysis is restricted to regions where an effective-mass plot
clearly shows that the ground state is dominant.

For the quantities discussed below, both jackknife and bootstrap
analyses of the correlators and effective masses ({\it e.g.},
$\log\left[G_n(t)/G_n(t-1)\right]$) are performed for each energy or
combination thereof. These are then used in correlated and
uncorrelated fits to the $t$ dependence to extract the relevant
quantity.  Our systematic uncertainties are determined by comparison
of our different analysis procedures and variation of the fit ranges.
To avoid uncertainties arising from scale setting, we focus on the
dimensionless quantities $m_\pi \overline{a}_{\pi^+\pi^+}$ and $m_\pi
f_\pi^4 \overline{\eta}_3^{L}$ ($\overline{\eta}_3^{L}$ is expected to
scale as $m_\pi^{-1} f_\pi^{-4}$ by NDA).

\begin{figure}[!ht]
  \centering
\vspace*{-4mm}
  \includegraphics[width=2.7in]{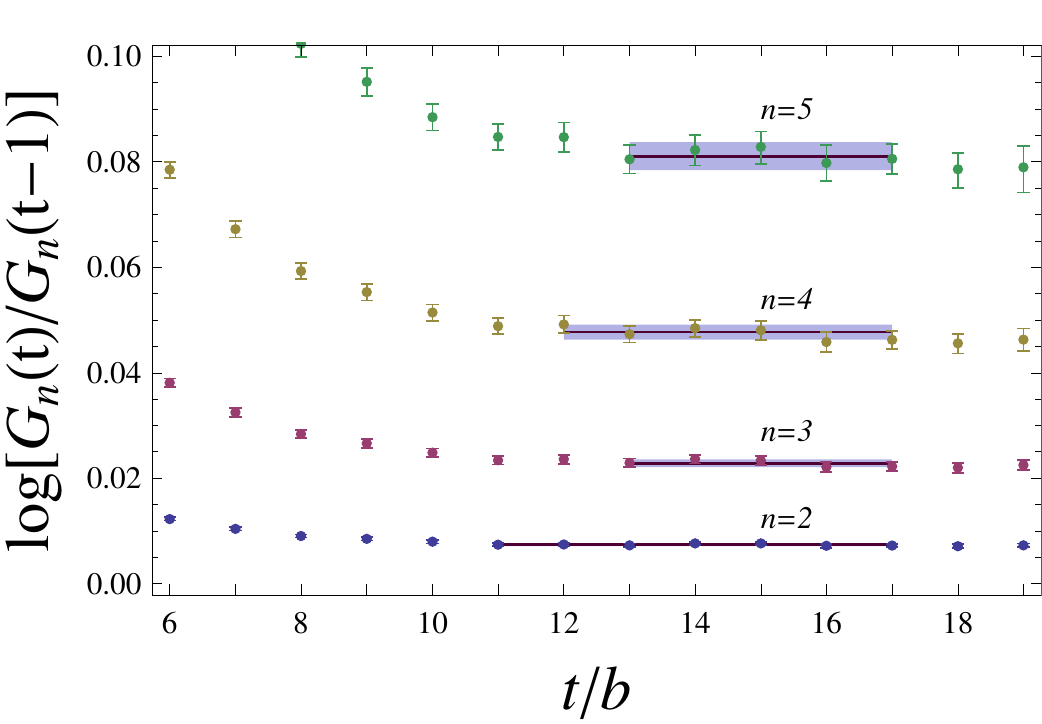}
\vspace*{-4mm}
  \caption{Energy shifts, $\Delta E_n$, for  $n=2,3,4,5$ $\pi^+$'s for
    the calculations with $m_\pi\sim 352~{\rm MeV}$.
The statistical and systematic uncertainties of the fits 
have been combined in quadrature.}
  \label{fig:En}
\end{figure}

\begin{figure}[!ht]
  \centering
\vspace*{-2mm}
  \includegraphics[width=2.8in]{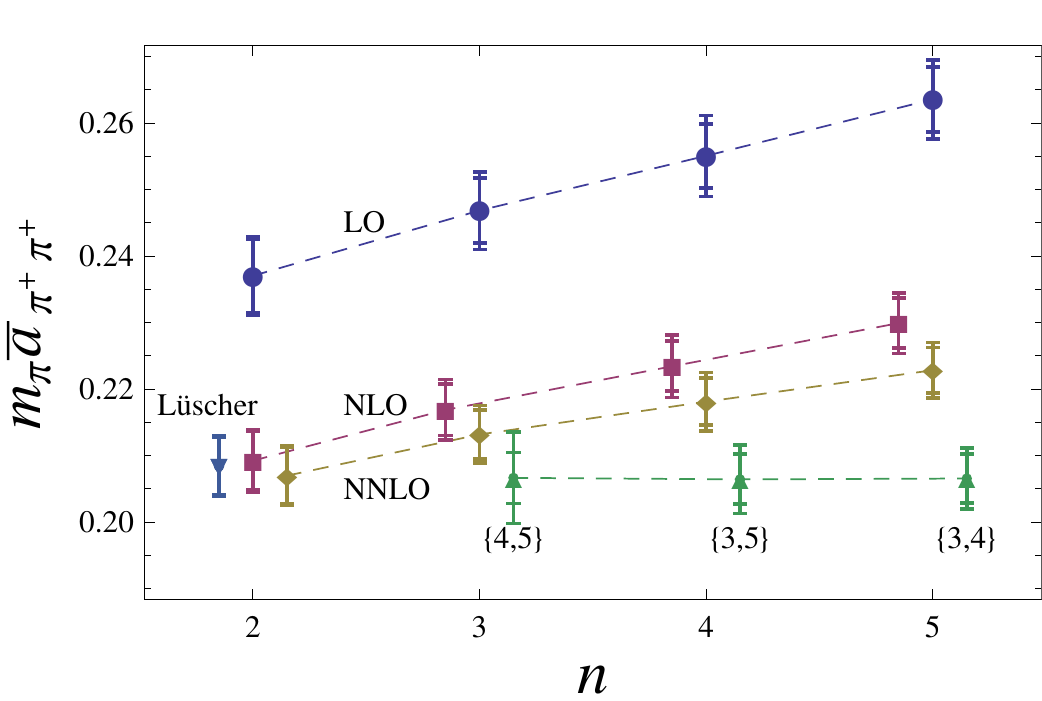}
\vspace*{-2mm}
  \caption{
    Extracted values of $m_\pi \overline{a}_{\pi^+\pi^+}$ at
    $m_\pi\sim 352~{\rm MeV}$.  LO, NLO and NNLO correspond to
    extractions of $\overline{a}$ at ${\cal O}(1/L^3,1/L^4,1/L^5)$
    from Eq. (\ref{eq:1}), respectively.  The N$^3$LO results for
    $\{n,m\}=\{3,4\}, \{3,5\}$, and $\{4,5\}$ are determined from Eq.
    (\ref{eq:2}).  For $n=2$, the exact solution of the eigenvalue
    equation \protect{\cite{Luscher:1986pf}} is denoted by
    ``L{\"u}scher''.}
  \label{fig:scatt}
\end{figure}

\begin{figure}[!ht]
  \centering
\vspace*{-2mm}
  \includegraphics[width=2.8in]{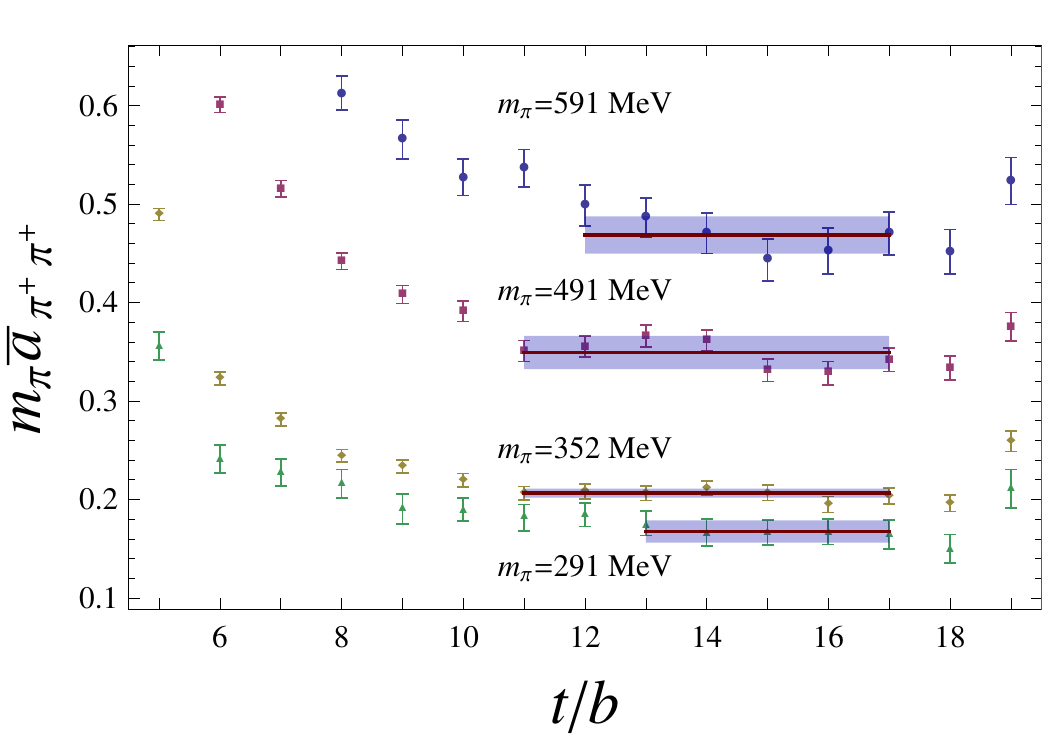}
\vspace*{-2mm}
  \caption{Effective $m_\pi \overline{a}_{\pi^+\pi^+}$ plot for
    $\{n,m\}=\{3,5\}$ using Eq. (\ref{eq:2}).  The statistical and
    systematic uncertainties of the fits have been combined in
    quadrature.  }
  \label{fig:mpia}
\end{figure}

The $\pi^+\pi^+$ scattering length (more precisely, the combination
$\overline{a}_{\pi^+\pi^+}$) has been studied repeatedly in lattice
QCD using the finite-volume formalism of L\"uscher
\cite{Luscher:1986pf} (for $a/L\ll1$ a perturbative expansion gives
the $n=2$ case of Eq.~(\ref{eq:1})).  In particular, a precise
extraction of this scattering length has been performed using the same
propagators as used in this work~\cite{Beane:2007xs}.  Therefore, the
utility of multi-pion energies in extracting the scattering length can
be ascertained, as summarized in Figs.~\ref{fig:En}, \ref{fig:scatt}
and \ref{fig:mpia}.

In Fig.~\ref{fig:En}, the energy shifts for $n=2$, 3, 4, and 5 are
displayed for the highest-precision calculation, $m_\pi\sim 352~{\rm
  MeV}$.  Clear plateaus are visible for each $n$; indeed, the
relative uncertainty decreases with increasing $n$ in the range
explored (this is particularly clear for the calculation with
$m_\pi\sim 291~{\rm MeV}$). Since multiple combinations of pions
interact in an $n$-pion state over a larger volume of the lattice, a
statistically-improved signal results.

Figure~\ref{fig:scatt} presents extractions of the scattering length
at all four orders in the $1/L$ expansion in Eq.~(\ref{eq:1}) for
$m_\pi\sim 352~{\rm MeV}$.  For $n>2$, the N$^3$LO ($1/L^6$)
extraction is performed using Eq.~(\ref{eq:2}) with the point at $n=3$
arising from the energy shifts $\Delta E_4$ and $\Delta E_5$, and so
on.  Significant dependence on $n$ is observed in the lower-order
extractions (LO, NLO and NNLO), indicating the presence of residual
finite-volume effects.  However the most accurate extractions using
Eq.~(\ref{eq:2}), which eliminates the three-$\pi^+$ interaction
(Eq.~(\ref{eq:1}) for $n=2$), are in close agreement for all $n$.
This provides a non-trivial check of the $n$-dependence of
Eq.~(\ref{eq:1}), particularly the presence of a term that scales as
{\tiny $\Choose{n}{3}$}, which can be identified as the three-pion
interaction.

The effective $m_\pi \overline{a}_{\pi^+\pi^+}$ plots for
$\{n,m\}=\{3,5\}$ are shown in Fig.~\ref{fig:mpia}.  Agreement is
found at the level of correlation functions with those of $n=2$,
$\{n,m\}=\{3,4\}$ and $\{n,m\}=\{4,5\}$.  This agreement suggests that
higher-order effects in $1/L$ (such as higher-derivative interactions
and four-particle interactions, which occur at ${\cal O}(1/L^8)$ and
${\cal O}(1/L^9)$, respectively) are small.  For the calculations with
$m_\pi\sim 291~{\rm MeV}$, the $n>3$ effective $m_\pi
\overline{a}_{\pi^+\pi^+}$ plots are significantly ``cleaner'' than
for $n=2$.

\begin{figure}[!ht]
  \centering
\includegraphics[width=2.8in]{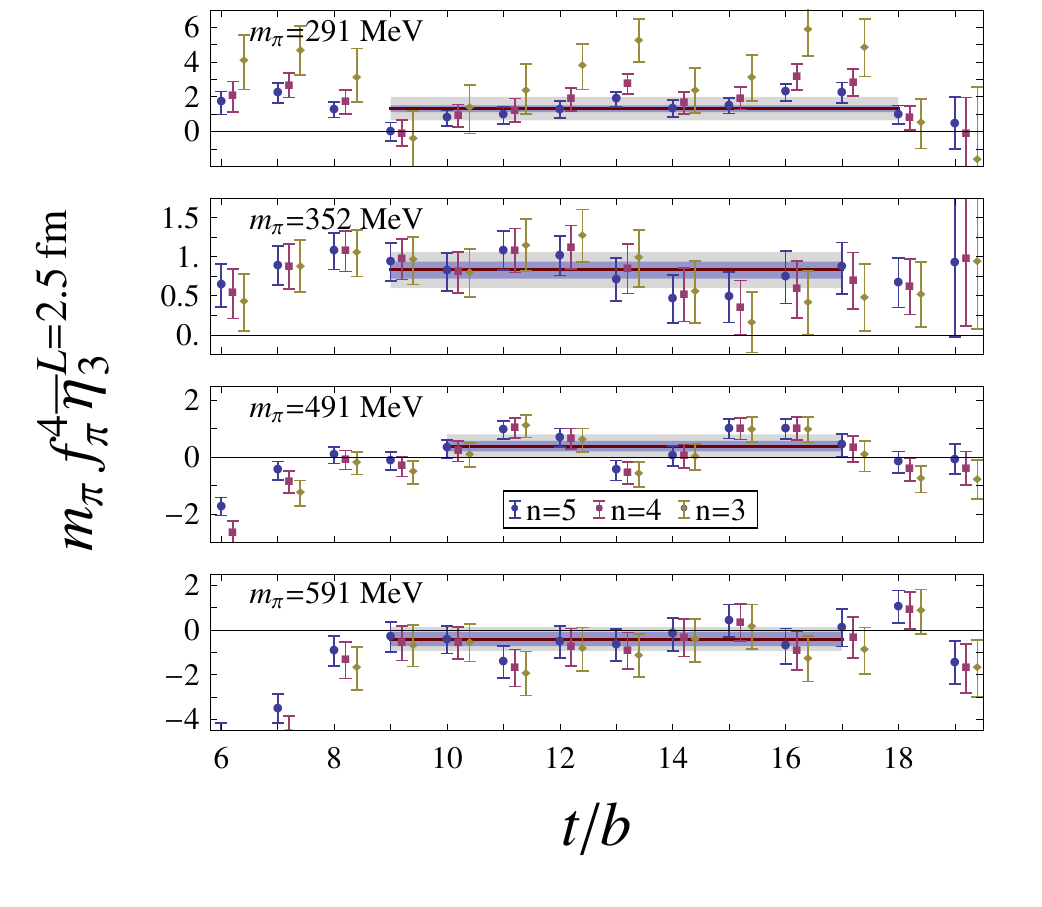}  
\vspace*{-4mm}
  \caption{Effective $m_\pi f_\pi^4  \overline{\eta}_3^L$  plots
    extracted from the $n=3$, 4, and 5 $\pi^+$ energy shifts. The fits
    shown correspond to the $n=5$ calculation.  The statistical and
    systematic uncertainties have been combined in quadrature.  }
  \label{fig:mpi5eta3bar}
\end{figure}

\begin{figure}[!ht]
\vspace*{-1mm}
  \centering
\includegraphics[width=2.7in]{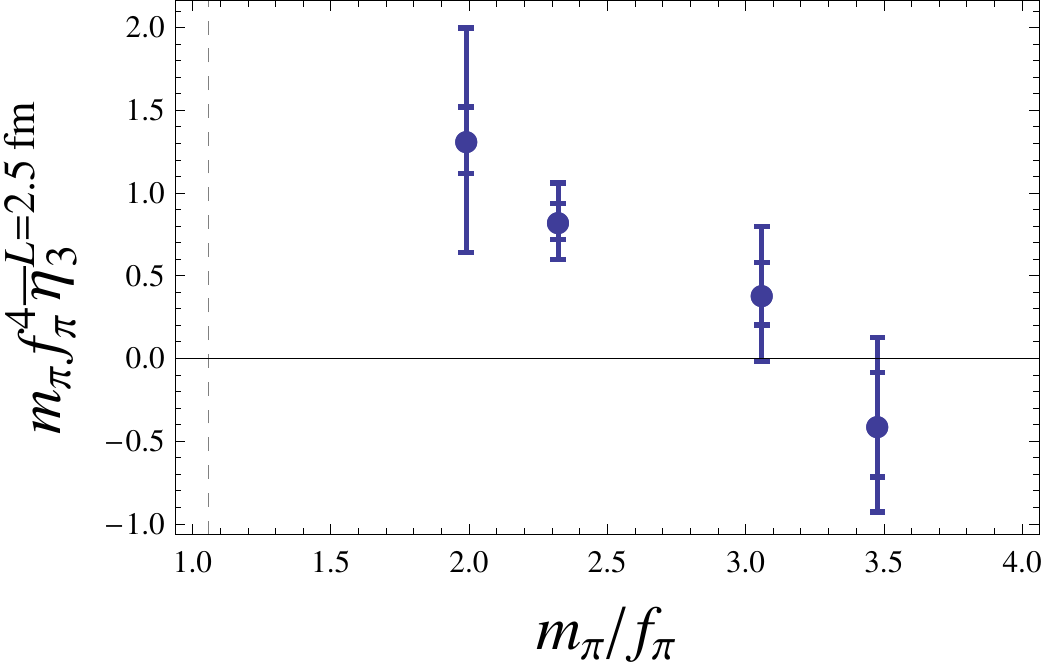}    
\vspace*{-2mm}
  \caption{Mass dependence of the RGI three-$\pi^+$ interaction 
    $m_\pi f_\pi^4 \overline{\eta}_3^L$.  The statistical and
    systematic uncertainties have been combined in quadrature.  The
    vertical dashed line denotes the physical value of
    $m_{\pi^+}/f_{\pi^+}$.  }
  \label{fig:mpieta3barvsmpi}
\end{figure}

To isolate the three-body interaction, we turn now to the combinations
defined in Eq.~(\ref{eq:3}), and the effective $m_\pi f_\pi^4
\overline{\eta}_3^L$ plots are shown in Fig.~\ref{fig:mpi5eta3bar}.  A
nonzero value of $m_\pi f_\pi^4 \overline{\eta}_3^L$ is found for
$m_\pi\sim 291$ and $352~{\rm MeV}$.
Figure~\ref{fig:mpieta3barvsmpi} and Table~\ref{tab:summ} summarize
the results for the RGI three-$\pi^+$ interaction, $m_\pi f_\pi^4
\overline{\eta}_3^L$, at $L=2.5$~fm.  The magnitude of the result is
consistent with NDA.  In Table \ref{tab:summ}, we also present $m_\pi
f_\pi^4 \eta_3(\mu=1/b)$, a quantity that has a well-defined
infinite-volume limit (unlike $\overline{\eta}_3^L$) but is scale and
scheme dependent.  Its scale dependence is given below
Eq.~(\ref{eq:1}), and we use the MS subtraction scheme~\cite{Beane:2007qr}.
\begin{table}[!h]
  \centering
  \begin{ruledtabular}
      \begin{tabular}{c|cccc}
     $m_\pi$ (MeV)   & $291$ & $352$ & $491$ & $591$
\\\hline
     $m_\pi f_\pi^4  \overline{\eta}_3^{L=2.5~{\rm fm}}$ & 1.3(2)(7) & 0.8(1)(2) & 0.4(2)(4)& -0.4(3)(4)  \\
$m_\pi f_\pi^4 \eta_3(\mu=1/b)$ & 1.2(2)(7) & 0.7(1)(2) & -0.1(2)(4) & -1.3(3)(4) 
      \end{tabular}
  \end{ruledtabular}
\caption{The $\pi^+\pi^+\pi^+$ interaction as defined in
  Eq.~(\protect\ref{eq:3}). The most precise result (using $n=5$) is quoted.
}
  \label{tab:summ}
\vspace*{-2mm}
\end{table}

Finally, Eq.~(\ref{eq:4}) and its counterparts involving other
combinations of energies allow for a determination of residual $1/L^7$
contributions to the quantities we have extracted at N$^3$LO.  They
are all consistent with zero and are $\lsim 0.05$.

Further lattice QCD calculations are required before a definitive
statement about the physical value of the three-pion interaction,
$m_\pi f_\pi^4 \overline{\eta}_3^L$, can be made.  While at lighter
pion masses, there is evidence for a contribution to the various
$n$-pion energies beyond two body scattering that scales as the
three-body contribution in Eq.~(\ref{eq:1}), a number of systematic
effects must be further investigated.  The extraction of this quantity
has corrections that are formally suppressed by $\overline{a}/L$,
however, the coefficient of the higher order term(s) may be large, and
the next order term in the volume expansion needs to be computed (for
$n=3$, this result is known \cite{Tan:2007bg}).  It is also possible
that the signals seen in Fig.~\ref{fig:mpi5eta3bar} are artifacts of
the lattice discretization, but the observed scaling that is
consistent with {\tiny $\Choose{n}{3}$} suggests this is not the case.
However, calculations at a finer lattice spacings and with different
lattice discretizations are required to resolve this issue.


As the lattice QCD study of nuclei is an underlying motivation for
this work, it is worth considering difficulties that will be
encountered in generalizing the result described here to baryonic
systems. Certain difficulties have been discussed in
Ref.~\cite{Beane:2007qr}. Here we focus on the numerical issues.  The
ratio of signal-to-noise scales very poorly for baryonic observables
\cite{Lepage:1989hd}, requiring an exponentially-large number of
configurations to extract a precise result.  Also, the factorial
growth of the combinatoric factors involved in forming the correlators
for large systems of bosons and fermions and the high powers to which
propagators are raised ({\it e.g.,} for the 12-$\pi^+$ correlator,
there is a term $43545600\ {\rm tr}[\Pi^{11}]{\rm tr}[\Pi]$) implies
that the propagators used to form the correlation functions must be
known to increasingly high precision.  There is much room for
theoretical advances in this area.

In this work we have numerically studied the ground-state energies of
$n=2,\,3,\,4,\,5$ $\pi^+$'s in a cubic volume with periodic boundary
conditions using lattice QCD.  We find that the $\pi^+\pi^+$
scattering length can be extracted from combinations of these energies
that eliminate the three-$\pi^+$ interaction, and agree with previous
$n=2$ calculations~\cite{Beane:2007xs}.  In some cases the precision
of the extraction is improved.  We have found evidence of a
repulsive three-$\pi^+$ interaction for $m_\pi\lsim 352~{\rm MeV}$.
Future calculations will extend these results to larger $n$ and to
systems involving multiple kaons and pions.  Further, calculations
must be performed in different spatial volumes to determine the
leading correction (${\cal O}(1/L)$) to the three-$\pi^+$ interaction,
and at different lattice spacings in order to eliminate finite-lattice
spacing effects, which are expected to be small.

\acknowledgements{We thank M.~Endres and D.~B.~Kaplan for discussions,
  A.~Parre\~no for propagator generation, and R.~Edwards and B.~Jo\'o
  for help with the Chroma/QDP++ software library \cite{chroma}. Our
  computations were performed at JLab, FNAL, LLNL, NCSA, and Centro
  Nacional de Supercomputaci\'on (Barcelona, Spain).  We acknowledge
  DOE Grants No.~DE-FG03-97ER4014 (MJS, WD), DE-AC05-06OR23177 (KO),
  DE-AC52-07NA27344 (TL) and W-7405-Eng-48 (TL), NSF CAREER Grant No.
  PHY-0645570 (SB, AT).  KO acknowledges the Jeffress Memorial Trust,
  grant J-813 and a DOE OJI grant DE-FG02-07ER41527.}

\end{document}